\documentclass{svproc}
\usepackage{url}

\usepackage{graphicx}
\usepackage{color}
\usepackage{cancel}
\usepackage{float}

\begin{document}
\mainmatter              
\title{$K^{*}(892)^0$ production in p+p interactions from NA61/SHINE}
\titlerunning{}  
%
\author{Angelika Tefelska for the NA61/SHINE Collaboration}
\authorrunning{Angelika Tefelska} 
%
\tocauthor{}
\institute{Warsaw University of Technology, Faculty of Physics\\
\email{angelika.tefelska@cern.ch}}

\maketitle              

\begin{abstract}

The measurement of $K^*(892)^0$ resonance production via its $K^+ \pi^-$ decay mode in inelastic p+p collisions  at  beam  momenta 40--158  GeV/c ($\sqrt{s_{NN}}$=8.8--17.3  GeV) is presented. The data were recorded by the NA61/SHINE hadron spectrometer at the CERN Super Proton Synchrotron. The analysis of $K^*(892)^0$ was done with the template method. The results include the double differential spectra $d^2 n/(dydp_T)$, $d^2 n/(m_T dm_T dy)$ as well as dn/dy spectra. 
\end{abstract}
\section{Introduction}

The study of short-lifetime resonances are unique tools to understand the less known aspects
of high energy collisions, especially its time evolution. The measurement of $K^{*}(892)^0$ meson production may help to distinguish between two possible scenarios for the fireball freeze-out: the sudden and the gradual one~\cite{markert}. The ratio of $K^{*}(892)^0$ to charged kaon production may allow to determine the time between chemical and kinetic freeze-outs \cite{markert,3}.

The transverse mass spectra and yields of $K^*(892)^0$ mesons are also important inputs for Blast-Wave models and Hadron Resonance Gas models. Moreover, resonance spectra and yields provide an important reference for tuning Monte Carlo string-hadronic models.

In this paper we report measurements of $K^{*}(892)^0$ resonance production via its $K^{+}\pi^{-}$ decay mode in inelastic p+p collisions at beam momenta 40--158~GeV/c
( $\sqrt{s_{NN}}$ = 8.8--17.3 GeV). The data were recorded by the NA61/SHINE hadron spectrometer~\cite{21} at the CERN SPS. The template fitting method was used to extract the $K^{*}(892)^0$ signal. This analysis method is also known as the cocktail fit method and was used by many other experiments such as ALICE, ATLAS, CDF, and CMS.

\section{Methodology}

The $K^{*}(892)^0$ analysis was done for p+p interactions based on data sets recorded in years 2009, 2010 and 2011 which contained about 5.26 x $10^6$ (beam momentum 40 GeV/c), 4.78 x $10^6$ (beam momentum 80 GeV/c) and 56.65 x $10^6$ (beam momentum 158 GeV/c) good-quality collisions of the proton beam with a 20 cm long liquid hydrogen target. The NA61/SHINE calibration, track and vertex reconstruction procedures and simulations are discussed in Refs.~\cite{22,23,26}.

The template method was used to extract raw $K^{*}(892)^0$ signals. In this method the background is described as a sum of two components: mixed events and Monte Carlo generated templates which describe the contribution of $K^+ \pi^-$ pairs coming from sources other than the $K^{*}(892)^0$. For the studied resonance in the small p+p system, the template method was found to be much more effective in estimating the background than the standard procedure relying on mixed events only. More information about the methodology is available in Ref.~\cite{CPOD}.

\section{Results}

\begin{figure}[H]
	\centering
	\includegraphics[width=0.7\textwidth]{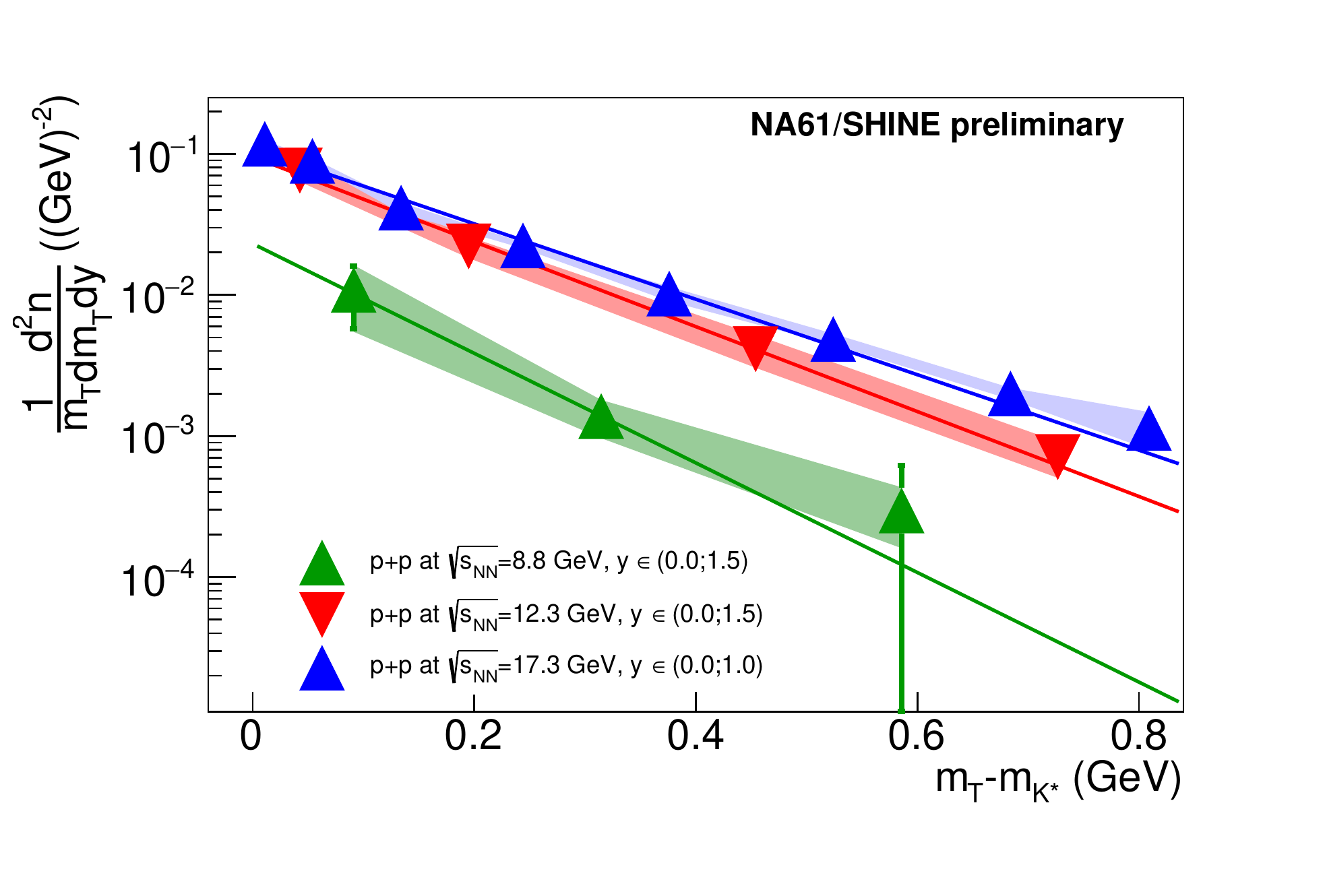}
	\caption{Preliminary results on mid-rapidity transverse mass spectra of $K^{*}(892)^0$ mesons produced in inelastic p+p collisions at 40, 80, and 158 GeV/c.}
	\label{fig:dndydmt}
\end{figure}

\begin{figure}
	\centering
	\includegraphics[width=0.7\textwidth]{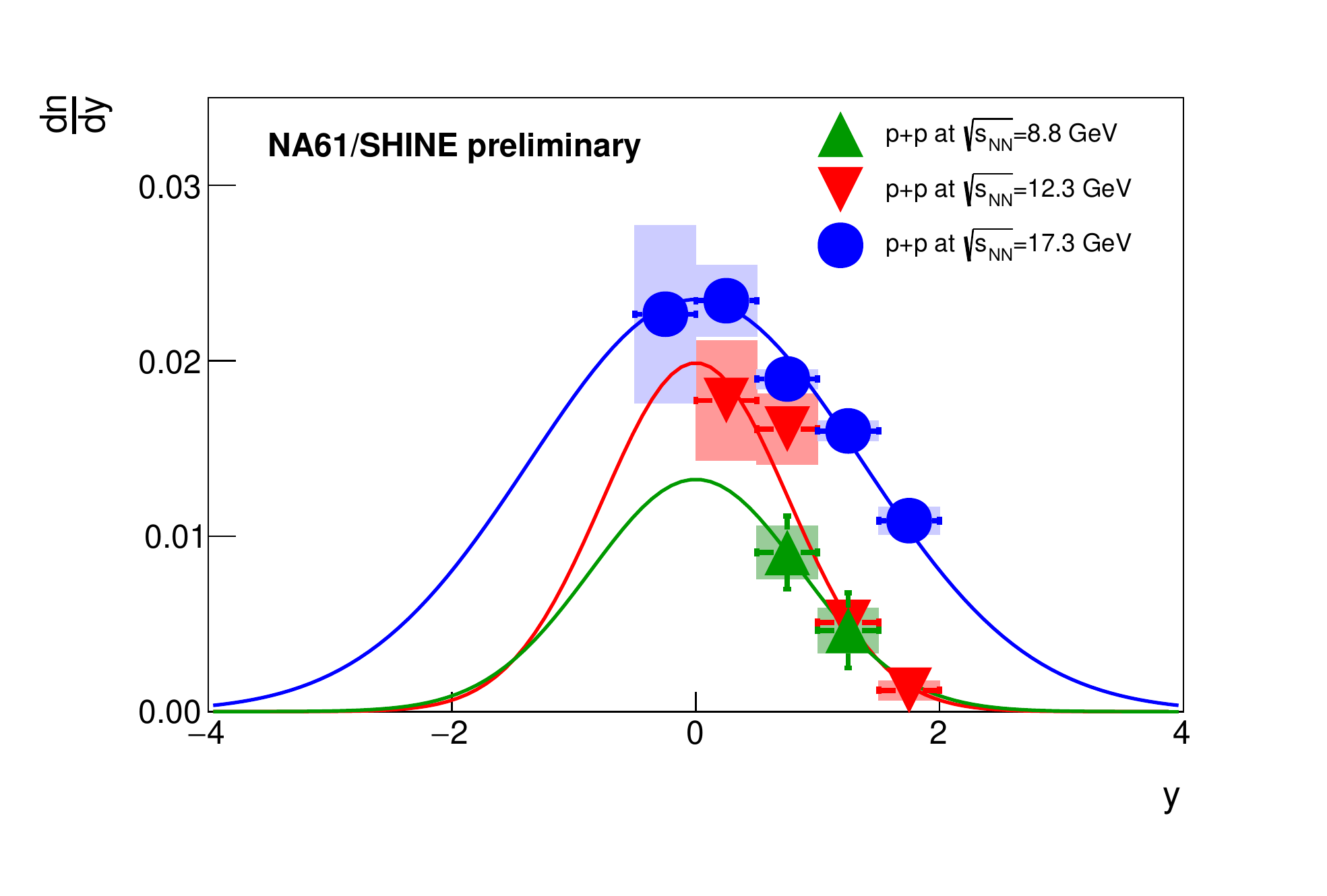}
	\caption{Preliminary results on rapidity spectra of $K^{*}(892)^0$ mesons produced in inelastic p+p
		collisions at 40, 80, and 158 GeV/c. For 40 and 80 GeV/c the results were obtained in a wide transverse momentum range (0 $<p_T<$ 1.5 GeV/c), whereas for 158 GeV/c $p_T$-extrapolated and integrated (0 $<p_T<$ $\infty$) results are shown.}
	\label{fig:dndy}
\end{figure}

Figure \ref{fig:dndydmt} presents the mid-rapidity transverse mass spectra of $K^{*}(892)^0$
mesons produced in inelastic p+p collisions at 40, 80, and 158 GeV/c. The rapidity spectra obtained by integrating and extrapolating the transverse momentum distributions are shown in Fig. \ref{fig:dndy}.

The mean multiplicities of $K^{*}(892)^0$ mesons in full phase-space (158 GeV/c) or in 0~$<p_T<$~1.5 GeV/c (40 and 80 GeV/c) were obtained from summing the measured points and adding the contribution from a Gaussian fit in the unmeasured region (lines in Fig. \ref{fig:dndy}). For 158 GeV/c the point with $y<0$ was calculated only to check the symmetry of the rapidity distribution and was not included in the procedure of mean multiplicity determination.  The numerical values of the mean multiplicities of $K^{*}(892)^0$ mesons are presented in Table \ref{tab:multiplicity}.

\begin{table}
	\centering
	\begin{tabular}{|c|c|c|}
		\hline
		$\sqrt{s_{NN}}$ & NA61/SHINE preliminary & NA49~\cite{NA49}\\
		\hline
		8.8 & 0.0285 $\pm$ 0.0031 $\pm$ 0.0046 & - \\
		\hline
		12.3 & 0.0381 $\pm$ 0.0054 $\pm$ 0.0037 & - \\
		\hline
		17.3 & 0.0806 $\pm$ 0.0006 $\pm$ 0.0026 & 0.0741 $\pm$ 0.0015 $\pm$ 0.0067 \\
		\hline
	\end{tabular}
	\vspace{0.5cm}
	\caption{Mean multiplicities of $K^{*}(892)^0$ mesons produced in inelastic p+p collisions at 40, 80, and 158 GeV/c from NA61/SHINE, as well as from NA49~\cite{NA49} at 158 GeV/c. The first uncertainty is statistical and the second is systematic.}
	\label{tab:multiplicity}
\end{table}

\section{Comparison with Hadron Resonance Gas Model}

\begin{figure}
	\centering
	\includegraphics[width=0.55\textwidth]{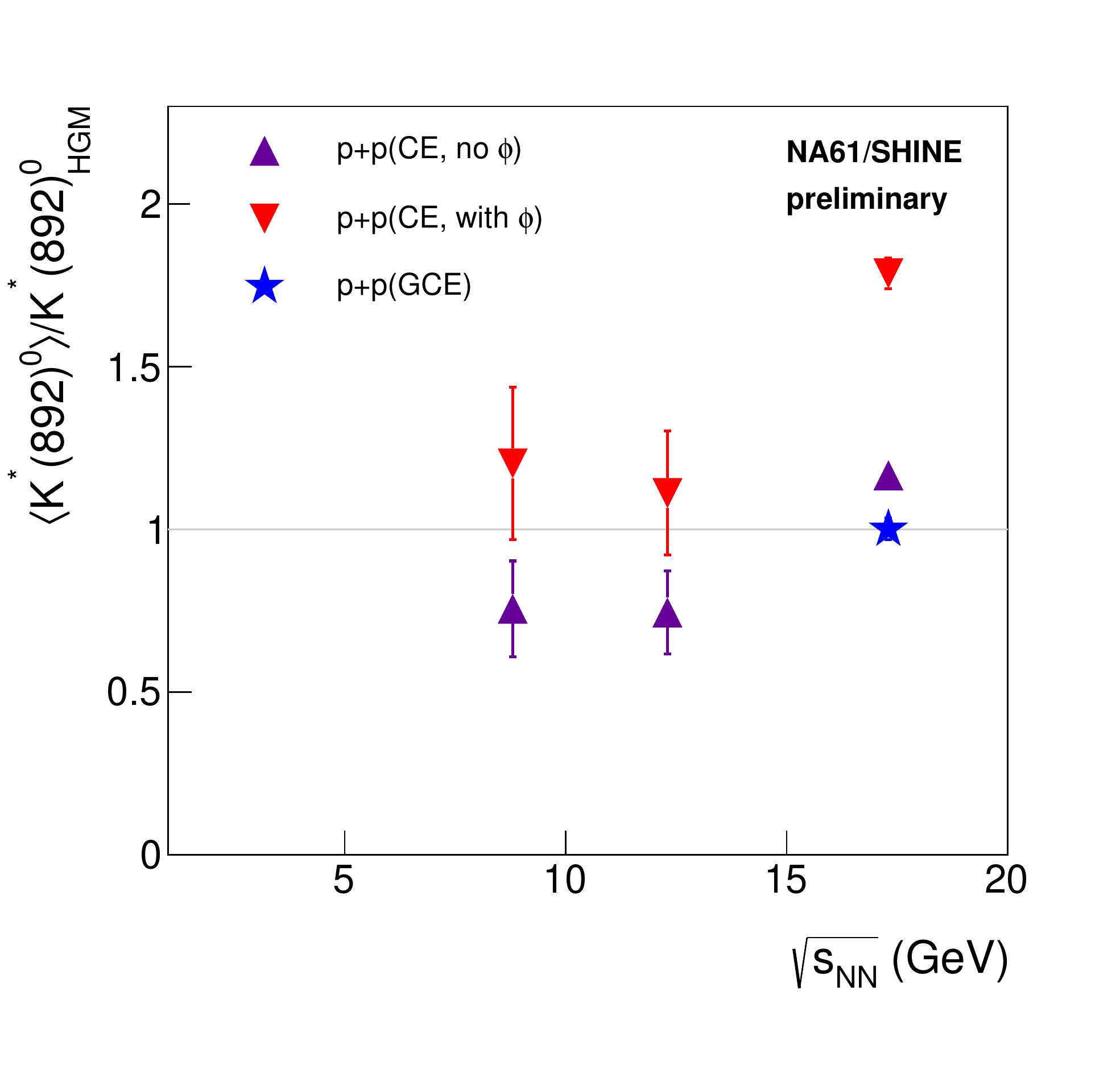} \caption{Preliminary results on mean multiplicities of $K^{*}(892)^0$ mesons produced in p+p collisions compared to Hadron Gas Model predictions~\cite{Begun1,Begun2}. Statistical and systematic uncertainties of $\langle K^{*}(892)^0 \rangle$ were added in quadrature.}
	\label{fig:HGM}
\end{figure}

The new $K^{*}(892)^0$ results of NA61/SHINE were compared with predictions of the statistical Hadron Resonance Gas model~\cite{Begun1,Begun2} in Canonical (CE) and Grand Canonical (GCE) formulations (Fig. \ref{fig:HGM}). At 158 GeV/c the GCE model provides a very good description of $K^{*}(892)^0$ production in the small p+p system. The CE model also agrees provided that the $\phi$ meson is excluded from the fits. More comparisons of the 158 GeV/c measurements with Hadron Resonance Gas model predictions of other authors can be found in Refs.~\cite{CPOD,CPOD_KG}.

\section{Acknowledgements}

This work was supported by the National Science Centre, Poland (grant \\ 2017/25/N/ST2/02575) and partially supported by the Ministry of Science and Higher Education, Poland (DIR/WK/2016/2017/10-1).

%
%

\end{document}